\documentclass[aps,prd,superscriptaddress,showpacs]{revtex4}
\usepackage{graphicx}
\usepackage{epstopdf}
\usepackage{amsmath}
\usepackage{amsfonts}
\usepackage{amssymb}
\usepackage{latexsym}

\begin{document}
\title{Aspects of screening and confinement in a topologically massive $U{\left( 1 \right)_{\cal W}} \times U{(1)_{\cal Y}}$ Chern-Simons-Higgs theory}
\author{Patricio Gaete} \email{patricio.gaete@usm.cl} 
\affiliation{Departamento de F\'{i}sica and Centro Cient\'{i}fico-Tecnol\'ogico de Valpara\'{i}so, Universidad T\'{e}cnica Federico Santa Mar\'{i}a, Valpara\'{i}so, Chile}
\author{Jos\'{e} A. Helay\"{e}l-Neto}\email{helayel@cbpf.br}
\affiliation{Centro Brasileiro de Pesquisas F\'{i}sicas (CBPF), Rio de Janeiro, RJ, Brasil} 
\date{\today}

\begin{abstract}
By using the gauge-invariant but path-dependent, variables formalism, we consider a recently proposed  topologically massive $U{\left( 1 \right)_{\cal W}} \times U{(1)_{\cal Y}}$ Chern-Simons-Higgs theory in $2+1$ dimensions. In particular, we inspect the impact of a Chern-Simons mixing term between two Abelian gauge fields on physical observables. We pursue our investigation by analysing the model in two different situations. In the first case, where we integrate out the massive excitation and consider an effective model for the massless field, we show that the interaction energy contains a linear term leading to the confinement of static charge probes along with a screening contribution. The second situation, where the massless field can be exactly integrated over with its constraint duly taken into account, the interesting feature is that the resulting effective model  describes a purely screening phase, without any trace of a confining regime.
\end{abstract}
\pacs{14.70.-e, 12.60.Cn, 13.40.Gp}
\maketitle

\section{Introduction}

Systems in $(2+1)$ dimensions and its physical consequences such as massive gauge fields and fractional statistics, where the physical excitations obeying it are called anyons, have been object of great interest for many authors \cite{Deser, Dunne,Khare,Banerjee}. As well known, three-dimensional Chern-Simons gauge theory is the key example so that Wilczek's charge-flux composite model of the anyon can be implemented \cite{Wilczek1,Wilczek2}. We further recall here that three-dimensional Yang-Mills theories are super-renormalizable and mass for the gauge fields are not in conflict with gauge symmetry \cite{Deser}. Interestingly, it has been shown that topologically massive Yang-Mills theories are ultraviolet finite \cite{Cima,Leme1,Leme2,Barnich}. Meanwhile, $(2+1)$-D theories may be adopted to describe the high-temperature limit of models in  $(3+1)$-D \cite{Das}. Incidentally, it is of interest to notice that planar gauge theories are useful to probe low-dimensional condensed matter systems, such as the description of boson collective excitations (like spin or pairing fluctuations) by means of effective gauge theories  and high-$T_{C}$ superconductivity, for which planarity is a very good approximation \cite{Khveschenko}. We also draw attention to the fact that $(2+1)$ D theories, specially Yang-Mills theories, are very important for a reliable comparison between results coming from the continuum and lattice calculations, for much larger lattices can be implemented in three space-time dimensions \cite{Feuchter}. More recently, $3D$ physics has been studied in connection to branes physics; for example, issues like self-duality \cite{Singh} and new possibilities for supersymmetry breaking as induced by $3$-branes \cite{Hanany,Klein1,Klein2} are of special relevance. Another interesting observation is that the quark-antiquark potential for some non-Abelian $(2+1)$-dimensional Yang-Mills theories has been studied in \cite{Orland1,Orland2,Orland3,Karabali1,Karabali2}.

We further note that recently a new approach to describe superconductivity at all temperatures has been considered \cite{Shapos1}. The crucial ingredient of this development is to introduce a Chern-Simons mixing term between two Abelian gauge fields, in other words, this new development is a topologically massive $U{\left( 1 \right)_{\cal W}} \times U{(1)_{\cal Y}}$ Chern-Simons-Higgs theory with a mixing term. More precisely, it was argued that by using a new basis, $\left\{ {{{\cal A}_\mu },{{\cal Z}_\mu }} \right\}$, and a specific condition between Chern-Simons coefficients must be satisfied, the model  can support a superconducting phase at all temperatures. Let us mention here that this new theory admits the existence of a new topological vortex solution.
It should be further noted that the Chern-Simons mixing term is the $(2+1)$-dimensional version of the BF theory \cite{Shapos2}. We also quote the recent work of \cite{Arias}. The authors build up vortex solutions for Abelian Chern-Simons-Higgs theories with visible and hidden sectors, where there also appear mixing terms as the ones of Ref. \cite{Shapos2}. An $N=2$-SUSY extension is worked out in Ref. 
\cite{Arias}.

Inspired by these observations, the purpose of this paper is to further elaborate on the physical content of topologically massive $U{\left( 1 \right)_{\cal W}} \times U{(1)_{\cal Y}}$ Chern-Simons-Higgs theory. 
Of particular concern to us is the effect of the new basis, $\left\{ {{{\cal A}_\mu },{{\cal Z}_\mu }} \right\}$, and the condition between Chern-Simon coefficients on a physical observable. To do this, we will work out the static potential for the theory under consideration by using the gauge-invariant but path-dependent variables formalism. According to this formalism, the interaction energy between two static charges is obtained once a judicious identification of the physical degrees of freedom is made \cite{Pato1,Pato2}. It also provides an alternative technique for determining the static potential for a gauge theory. When we compute in this way the static potential, the result of this calculation is rather unexpected in the case of an effective Lagrangian in terms of the $\cal A$ field. It is shown that the interaction energy displays a screening part, encoded by Bessel functions, and a linear confining potential. Incidentally, the above static potential profile is similar to that encountered in both Maxwell-like three-dimensional models induced by the condensation of topological defects \cite{Pato3} and by the condensation of charged scalars in $D=3$ dimensions \cite{Pato4}. In this way, we may establish a new connection among diverse models as well as exploiting this equivalence in explicit calculations. On the other hand, in the case of an effective Lagrangian in terms of the $\cal Z$ field, the static potential remains in a screening phase. We further note that related models were discussed in \cite{Jackiw,Rocek,Fradkin}. In particular, in \cite{Fradkin} the method of integrating one or the other gauge field in models with Maxwell and Chern-Simons terms has been employed to establish the duality between topologically massive and self-dual gauge theories first discussed in \cite{Jackiw}. Moreover, the non-Abelian extension of such kind of analysis has also been developed in \cite{Manias}.

Our work is organized according to the following outline: in Section II, we introduce the model and analyze the condition between Chern-Simons coefficients. In Section III, we compute the interaction energy for both effective Lagrangians. Finally, some concluding remarks are made in Sec. IV. In Appendix A, we collect some constants appearing in the static potential profile.

\section{Three-dimensional  Chern-Simons mixing terms} 

As mentioned above, the gauge theory we are considering is a recently proposed topological massive $U{\left( 1 \right)_{\cal W}} \times U{(1)_{\cal Y}}$ Chern-Simons-Higgs theory \cite{Shapos1}. The model is described by the three-dimensional space-time Lagrangian density:
\begin{equation}
{\cal L} =  - \frac{1}{4}{{\cal Y}_{\mu \nu }}{{\cal Y}^{\mu \nu }} - \frac{1}{4}{{\cal W}_{\mu \nu }}{{\cal W}^{\mu \nu }} + {\mu _{\cal Y}}{\varepsilon ^{\mu \nu \alpha }}{{\cal Y}_{\mu \nu }}{{\cal Y}_\alpha } - {\mu _{\cal W}}{\varepsilon ^{\mu \nu \alpha }}{{\cal W}_{\mu \nu }}{{\cal W}_\alpha } + |{D_\mu }\phi{|^2} - V\left( {\phi {\phi ^ * }} \right), \label{CSm05}
\end{equation}
where 
\begin{equation}
V\left( {\phi ,{\phi ^ * }} \right) = {m^2}\phi {\phi ^ * } + \frac{\lambda }{4}|\phi {\phi ^ * }{|^2}. \label{CSm10}
\end{equation}
Here ${{\cal Y}_{\mu \nu }} = {\partial _\mu }{{\cal Y}_\nu } - {\partial _\nu }{{\cal Y}_\mu }$, ${{\cal W}_{\mu \nu }} = {\partial _\mu }{{\cal W}_\nu } - {\partial _\nu }{{\cal W}_\mu }$, with ${D_\mu } = {\partial _\mu } - i{g_1}{{\cal Y}_\mu } - i{g_2}{{\cal W}_\mu }$, representing the covariant derivative. In passing we note that the mass square parameter $m^2$ can be taken to be positive or negative. We also point out that the Chern Simons coefficients ${\mu _{\cal Y}}$, ${\mu _{\cal W}}$, the mass parameter $m$ and the Higgs self coupling $\lambda$ have mass dimension $M$, whereas the coupling constants $g_1$ and $g_2$ have mass dimension ${M^{{\raise0.5ex\hbox{$\scriptstyle 1$}
\kern-0.1em/\kern-0.15em\lower0.25ex\hbox{$\scriptstyle 2$}}}}$.

As calculated in the work of Ref. \cite{Shapos1}, in the phase the Higgs scalar acquires a non-trivial V.E.V. , the spectrum exhibits massive excitations corresponding to the ${{\cal Y}_{\mu}}$- and ${{\cal Z}_{\mu}}$- fields. Nevertheless, in the particular case the ${\mu _{\cal Y}}$- and ${\mu _{\cal W}}$-parameters obey the relationship
\begin{equation}
{\mu _{\cal Y}} = {\mu _{\cal W}}{\tan ^2}\theta,   \label{CSm15}
\end{equation}
where $\tan \theta  = \frac{{{g_1}}}{{{g_2}}}$, one of the vector excitations become massless, namely,
\begin{equation}
{{\cal A}_\mu } \equiv \cos \theta {{\cal Y}_\mu } - \sin \theta {{\cal W}_\mu }, \label{CSm20}
\end{equation}
whereas the orthogonal field combination
\begin{equation}
{{\cal Z}_\mu } \equiv \sin \theta {{\cal Y}_\mu } + \cos \theta {{\cal W}_\mu }, \label{CSm25}
\end{equation}
exhibits the mass ${m_{\cal Z}}$. This is the regime of system we consider from now.

In such a situation, and considering we have chosen to work in the unitary gauge, where the phase of the 
$\phi$-field is gauged away, we shall have:
\begin{equation}
{{\cal F}_{\mu \nu }} = {\partial _\mu }{{\cal A}_\nu } - {\partial _\nu }{{\cal A}_\mu }, \label{CSm30}
\end{equation}
and
\begin{equation}
{{\cal Z}_{\mu \nu }} = {\partial _\mu }{{\cal Z}_\nu } - {\partial _\nu }{{\cal Z}_\mu }. \label{CSm35}
\end{equation}
The covariant derivative on the Higgs scalar turns out to be given by ${D_\mu } = {\partial _\mu } - ie{{\cal Z}_\mu }$ and $e \equiv \sqrt {g_1^2 + g_2^2}$.
The Lagrangian density, in such a case, reads as below:
\begin{equation}
{\cal L} =  - \frac{1}{4}{{\cal F}_{\mu \nu }}{{\cal F}^{\mu \nu }} - \frac{1}{4}{{\cal Z}_{\mu \nu }}{{\cal Z}^{\mu \nu }} + {\mu _1}{\varepsilon ^{\mu \nu \alpha }}{{\cal F}_{\mu \nu }}{{\cal Z}_\alpha } + \frac{{{\mu _2}}}{2}{\varepsilon ^{\mu \nu \alpha }}{{\cal Z}_{\mu \nu }}{{\cal Z}_\alpha } + \frac{1}{2}|{D_\mu }\phi {|^2} - V\left( {\phi ,{\phi ^ * }} \right), \label{CSm45}
\end{equation}
where the ${\mu _1}$- and ${\mu _2}$-parameters are given by 
\begin{equation}
{\mu _1} = 2{\mu _{\cal W}}\tan \theta, \label{CSm50}
\end{equation}
and
\begin{equation}
{\mu _2} = 2{\mu _{\cal W}}({\tan ^2}\theta  - 1). \label{CSm55}
\end{equation}
It is remarkable to notice that, despite spontaneous symmetry breaking has taken place, and the Higgs scalar is charged under both the $U(1)$-factors, only one vector boson, ${\cal Z^{\mu}}$, becomes massive; clearly, that is possible whenever ${\mu _{\cal Y}} = {\mu _{\cal W}}{\tan ^2}\theta$.
The ${\cal A}{\mu}$-field contribution is massless and the ${\cal Z}{\mu}$-field mass ($m_{\cal Z}$) has contributions coming from the ${\mu _1}$-, the ${\mu _2}$- and $|{D_\mu }\phi {|^2}$-terms of the Lagrangian (\ref{CSm45}), where we assume the splitting
\begin{equation}
\phi  = {\phi ^ * } = {\phi _0} + \frac{1}{{\sqrt 2 }}\eta \left( x \right), \label{CSm60}
\end{equation}
where ${\phi _0} = \sqrt {\frac{2}{\lambda }} |m|$, as a result of the choice of unitary gauge.

Inspired by these observations, the purpose of this work is to further elaborate on the physical content of this new topological massive $U{\left( 1 \right)_{\cal A}} \times U{(1)_{\cal Z}}$ Chern-Simons-Higgs theory.

\section{Interaction energy}

\subsection{Chern-Simons-Higgs effective theory I (integrating out the ${\cal Z}_\mu$-field)}

We turn our attention to the calculation of the interaction energy between static point-like sources for this theory by using the gauge-invariant but path-dependent variables formalism. However, before proceeding with the determination of the interaction energy, we first note that the Lagrangian density  (\ref{CSm45}) may be written as 
\begin{eqnarray}
{\cal L} &=&  - \frac{1}{4}{{\cal F}_{\mu \nu }}{{\cal F}^{\mu \nu }} + \frac{1}{2}{{\cal Z}^\mu }\left[ {\left( {\Delta  + {e^2}\phi {\phi ^ * }} \right){\eta _{\mu \nu }} - {\partial _\mu }{\partial _\nu } - 2{\mu _2}{\varepsilon _{\mu \nu \alpha }}{\partial ^\alpha }} \right]{{\cal Z}^\nu } 
+ \left( {\frac{e}{2}J_s^\alpha  + {\mu _1}{\varepsilon ^{\mu \nu \alpha }}{{\cal F}_{\mu \nu }}} \right){{\cal Z}_\alpha } \nonumber\\
&+& \frac{1}{2}{\partial ^\mu }{\phi ^ * }{\partial _\mu }\phi  - {m^2}{\phi ^ * }\phi  - \frac{\lambda }{4}|{\phi ^ * }\phi {|^2}, \label{Sener05}
\end{eqnarray}
where $J_s^\mu  \equiv i\left( {{\phi ^ * }{\partial ^\mu }\phi  - \phi {\partial ^\mu }{\phi ^ * }} \right)$ and $\Delta  \equiv {\partial _\mu }{\partial ^\mu }$.

As already expressed, by splitting the $\phi$-field into a background value and a dynamical part
\begin{equation}
\phi  = {\phi ^ * } = {\phi _0} + \frac{1}{{\sqrt 2 }}\eta \left( x \right), \label{Sener05b}
\end{equation}
where ${\phi _0} = \sqrt {\frac{2}{\lambda }} |m|$, we expand the Lagrangian up to quadratic terms in the fluctuations. Accordingly, equation (\ref{Sener05}) becomes
\begin{equation}
{\cal L} =  - \frac{1}{4}{{\cal F}_{\mu \nu }}{{\cal F}^{\mu \nu }} + \frac{1}{2}{{\cal Z}^\mu }\left[ {\left( {\Delta  + m_{\cal F}^2} \right){\eta _{\mu \nu }} - {\partial _\mu }{\partial _\nu } - 2{\mu _2}{\varepsilon _{\mu \nu \alpha }}{\partial ^\alpha }} \right]{{\cal Z}^\nu } + {\mu _1}{\varepsilon ^{\mu \nu \alpha }}{{\cal F}_{\mu \nu }}{{\cal Z}_\alpha }, \label{Sener10}
\end{equation}
where $m_{\cal F}^2 \equiv {e^2}\phi _0^2$. Next, by integrating out the $\cal Z$-field induces an effective theory for the $\cal A$-field. This leads us to the following effective Lagrangian density:
\begin{equation}
{\cal L} =  - \frac{1}{4}{F_{\mu \nu }}\left[ {\frac{{P{\Delta ^2} + Q\Delta  + R}}{{{\Delta ^2} + D\Delta  + m_{\cal F}^4}}} \right]{F^{\mu \nu }} - \mu _1^2s{\varepsilon ^{\nu \rho \lambda }}{F_{\nu \kappa }}\frac{{{\partial ^\kappa }}}{{\left( {{\Delta ^2} + D\Delta  + m_{\cal F}^4} \right)}}{F_{\rho \lambda }}-A_0J^0, \label{Sener15}
\end{equation}
where $s = 2{\mu _2}$ and $J^0$ is an external current. Whereas $P = \left( {1 + \frac{{2\mu _1^2}}{{m_{\cal F}^2}}} \right)$, $Q = 6\mu _1^2 + 2m_{\cal F}^2 + {s^2} + \frac{{2\mu _1^2{s^2}}}{{m_{\cal F}^2}}$, $R = m_{\cal F}^2(m_{\cal F}^2 + 4\mu _1^2)$ and $D = 2m_{\cal F}^2 + {s^2}$.

Before going ahead, it should be noted that the theory described by equation (\ref{Sener15}) contains higher time derivatives, hence to construct the Hamiltonian special care has to be exercised. It should, however, be emphasized here that the present paper is aimed at studying the static potential of the above theory, hence in what follows we shall replace $\Delta$ by $-{\nabla ^2}$. Thus, the canonical quantization of this theory from the Hamiltonian point of view follows straightforwardly, as we shall show it below.

Having established the new effective Lagrangian, we can now compute the interaction energy. To this end, we first consider the Hamiltonian structure of the theory under consideration. The canonical momenta read
\begin{equation}
{\Pi ^\mu } =  - \left( {\frac{{P{\nabla ^4} - Q{\nabla ^2} + R}}{{{\nabla ^4} - D{\nabla ^2} + m_{\cal F}^4}}} \right){F^{0\mu }} - 12\mu _1^2s\frac{1}{{\left( {{\nabla ^4} - D{\nabla ^2} + m_{\cal F}^4} \right)}}{\partial ^\mu }B.             \label{Sener20}
\end{equation}
It is easy to see that $\Pi ^0$ vanishes, we then have the usual constraint equation, which according to Dirac's theory is written as a weak ($\approx$) equation, that is, $\Pi ^0\approx0$. It may be noted that the remaining non-zero momenta must also be written as weak equations. In such a case, ${\Pi _i} \approx \left( {\frac{{P{\nabla ^4} - Q{\nabla ^2} + R}}{{{\nabla ^4} - D{\nabla ^2} + m_{\cal F}^4}}} \right){E_i} - 12\mu _1^2s\frac{1}{{\left( {{\nabla ^4} - D{\nabla ^2} + m_{\cal F}^4} \right)}}{\partial _i}B$. The canonical Hamiltonian $H_C$ is then
\begin{eqnarray}
{H_C} &\approx& \int {{d^2}x} \left\{ {{\Pi ^i}{\partial _i}{A_0} + \frac{1}{2}{E_i}\left( {\frac{{P{\nabla ^4} - Q{\nabla ^2} + R}}{{{\nabla ^4} - D{\nabla ^2} + m_{\cal F}^4}}} \right){E_i} + \frac{1}{2}B\left( {\frac{{P{\nabla ^4} - Q{\nabla ^2} + R}}{{{\nabla ^4} - D{\nabla ^2} + m_{\cal F}^4}}} \right)B} \right\} \nonumber\\
&-& 6\mu _1^2s\int {{d^2}x} {E_i}\frac{1}{{\left( {{\nabla ^4} - D{\nabla ^2} + m_{\cal F}^4} \right)}}{\partial _i}B+A_0J^0,\label{Sener25}
\end{eqnarray}
which must also be written as a weak equation. Next, the primary constraint, $\Pi ^0\approx0$, must be satisfied for all times. Accordingly, by using the equation of motion, $\dot Z \approx \left[ {Z,{H_C}} \right]$, we obtain the secondary constraint (Gauss's law) $\Gamma _1  \equiv \partial _i \Pi ^i -J^0 \approx 0$, which must also be true for all time. It can be easily seen that the stability of this constraint does not generate further constraints.
Hence, there are two constraints, which are first class. According to the general theory we obtain the extended Hamiltonian as an ordinary (or strong) equation by adding all the first-class constraints with arbitrary constraints. We thus write $H = H_C  + \int {d^3 x} \left( {u_0(x) \Pi_0(x)  + u_1(x) \Gamma _1(x) } \right)$, where $u_o(x)$ and $u_1(x)$ are arbitrary Lagrange multipliers. It is also important to observe that when this new Hamiltonian is employed, the equation of motion of a dynamic variable may be written as a strong equation. It should be further noted that $\dot A_0 \left( x \right) = \left[ {A_0 \left( x \right),H} \right] = u_0 \left( x \right)$, which is an arbitrary function. Since $\Pi^0\approx0$ always, neither $A^0$ nor $\Pi^0$ are of interest in describing the system and may be discarded from the theory. In fact, the term containing $A_0$ is redundant, because it can be absorbed by redefining the function $w(x)$. The Hamiltonian is thus given by 
\begin{eqnarray}
{H} &=& \int {{d^2}x} \left\{ {w(x)({\partial _i}{\Pi ^i}-J^0) + \frac{1}{2}{E_i}\left( {\frac{{P{\nabla ^4} - Q{\nabla ^2} + R}}{{{\nabla ^4} - D{\nabla ^2} + m_{\cal F}^4}}} \right){E_i} + \frac{1}{2}B\left( {\frac{{P{\nabla ^4} - Q{\nabla ^2} + R}}{{{\nabla ^4} - D{\nabla ^2} + m_{\cal F}^4}}} \right)B} \right\} \nonumber\\
 &-& 6\mu _1^2s\int {{d^2}x} {E_i}\frac{1}{{\left( {{\nabla ^4} - D{\nabla ^2} + m_{\cal F}^4} \right)}}{\partial _i}B,  \label{Sener30}
\end{eqnarray}
where $w(x) = u_1 (x) - A_0 (x)$.

Now the presence of the new arbitrary function, $w(x)$, is undesirable since we have no way of giving it a meaning in a quantum theory. A way to avoid this difficulty is to introduce a gauge condition such that the full set of constraints becomes second class. A particularly convenient gauge-fixing condition is
\begin{equation}
\Gamma _2 \left( x \right) \equiv \int\limits_{C_{\xi x} } {dz^\nu
} A_\nu \left( z \right) \equiv \int\limits_0^1 {d\lambda x^i }
A_i \left( {\lambda x} \right) = 0. \label{Sener35}
\end{equation}
where  $\lambda$ $(0\leq \lambda\leq1)$ is the parameter describing
the space-like straight path $ x^i = \xi ^i  + \lambda \left( {x -
\xi } \right)^i $, and $ \xi $ is a fixed point (reference point).
There is no essential loss of generality if we restrict our
considerations to $ \xi ^i=0 $. With this, we arrive at the only non-vanishing equal-time Dirac bracket for the canonical variables
\begin{equation}
\left\{ {A_i \left( {\bf x} \right),\Pi ^j \left( {\bf y} \right)} \right\}^ *
= \delta _i^j \delta ^{\left( 2 \right)} \left( {{\bf x} - {\bf y}} \right) -
\partial _i^x \int\limits_0^1 {d\lambda x^i } \delta ^{\left( 2
\right)} \left( {\lambda {\bf x} - {\bf y}} \right). \label{Sener40}
\end{equation}

Similarly, we write the Dirac brackets in terms of the magnetic ($B = {\varepsilon _{ij}}{\partial ^i}{A^j}$) and
electric (${E_i} = \left( {\frac{{{\nabla ^4} - D{\nabla ^2} + m_{\cal F}^4}}{{P{\nabla ^4} - Q{\nabla ^2} + R}}} \right){\Pi _i} + 12\mu _1^2s\frac{1}{{\left( {P{\nabla ^4} - Q{\nabla ^2} + R} \right)}}{\partial _i}B$) fields as
\begin{equation}
{\left\{ {{E_i}(x),B(y)} \right\}^ * } =  - \left( {\frac{{{\nabla ^4} - D{\nabla ^2} + m_{\cal F}^4}}{{P{\nabla ^4} - Q{\nabla ^2} + R}}} \right){\varepsilon _{ij}}\partial _x^j{\delta ^{\left( 2 \right)}}\left( {x - y} \right), \label{Sener45}
\end{equation}

\begin{equation}
{\left\{ {B\left( x \right),B\left( y \right)} \right\}^ * } = 0, \label{Sener50}
\end{equation}

\begin{eqnarray}
{\left\{ {{E_i}(x),{E_j}(y)} \right\}^ * } = 12\mu _1^2s\frac{{({\nabla ^4} - D{\nabla ^2} + m_{\cal F}^4)}}{{{{(P{\nabla ^4} - Q{\nabla ^2} + R)}^2}}}({\varepsilon _{jk}}\partial _i^x - {\varepsilon _{ik}}\partial _j^x)\partial _k^x{\delta ^{\left( 2 \right)}}\left( {x - y} \right). \label{Sener55}
\end{eqnarray}

One can now easily derive the equations of motion for the electric and magnetics fields. We find
\begin{equation}
\dot B\left( x \right) =  - {\varepsilon _{ij}}{\partial _i}{E_j}\left( x \right), \label{Sener60}
\end{equation}
and
\begin{equation}
{\dot E_i}\left( x \right) =  - 6\mu _1^2s\frac{1}{{\left( {P{\nabla ^4} - Q{\nabla ^2} + R} \right)}}{\varepsilon _{ij}}{\partial _j}{\partial _k}{E_k} + \left[ {1 + 72\mu _1^2s\frac{{{\nabla ^2}}}{{{{\left( {P{\nabla ^4} - Q{\nabla ^2} + R} \right)}^2}}}} \right]{\varepsilon _{ij}}{\partial _j}B. \label{Sener65}
\end{equation}
In the same way, we write Gauss's law as
\begin{equation}
\frac{{\left( {P{\nabla ^4} - Q{\nabla ^2} + R} \right)}}{{\left( {{\nabla ^4} - D{\nabla ^2} + m_{\cal F}^4} \right)}}{\partial _i}{E_i} - 12\mu _1^2s\frac{{{\nabla ^2}}}{{\left( {{\nabla ^4} - D{\nabla ^2} + m_{\cal F}^4} \right)}}B = ( - {J^0}).\label{Sener70}
\end{equation}

It is clear that, under the assumed conditions of static fields, equations (\ref{Sener60}) and (\ref{Sener65}) must vanish. In this manner, we obtain that the static electric field is given by
\begin{equation}
{E_i} = {\partial _i}\left[ {\frac{1}{{{\nabla ^2}}}\frac{{\left( {{\nabla ^4} - D{\nabla ^2} + m_{\cal F}^4} \right)}}{{\left( {P{\nabla ^4} - Q{\nabla ^2} + R} \right)}}} \right]\left( { - {J^0}} \right) + 72\mu _1^4s{\partial _i}\left[ {\frac{{\left( {{\nabla ^4} - D{\nabla ^2} + m_{\cal F}^4} \right)}}{{{{\left( {P{\nabla ^4} - Q{\nabla ^2} + R} \right)}^3}}}} \right]\left( { - {J^0}} \right).  \label{Sener75}
\end{equation}

After some further manipulations, the foregoing equation can be brought to the form 
\begin{eqnarray}
{E_i}(x) &=& \frac{1}{{\sqrt {{Q^2} - 4PR} }}{\partial _i}\left\{ {\left[ {\frac{{{\nabla ^2}}}{{{\nabla ^2} - M_1^2}} - \frac{{{\nabla ^2}}}{{{\nabla ^2} - M_2^2}}} \right] - D\left[ {\frac{1}{{{\nabla ^2} - M_1^2}} - \frac{1}{{{\nabla ^2} - M_2^2}}} \right]} \right\}\left( { - {J^0}} \right) \nonumber\\
&+& \frac{{m_{\cal F}^4}}{{\sqrt {{Q^2} - 4PR} }}{\partial _i}
\left[ {\frac{1}{{{\nabla ^2}\left( {{\nabla ^2} - M_1^2} \right)}} - \frac{1}{{{\nabla ^2}\left( {{\nabla ^2} - M_2^2} \right)}}} \right]\left( { - {J^0}} \right) \nonumber\\
&+&72\mu _1^4{s^2}{\partial _i}\left\{ {\left[ {\frac{{{A_1}}}{{{{\left( {{\nabla ^2} - M_1^2} \right)}^3}}} - {A_2}\frac{{{\nabla ^2}}}{{{{\left( {{\nabla ^2} - M_1^2} \right)}^3}}}} \right] - \left[ {\frac{{{A_3}}}{{{{\left( {{\nabla ^2} - M_2^2} \right)}^3}}} - {A_4}\frac{{{\nabla ^2}}}{{{{\left( {{\nabla ^2} - M_2^2} \right)}^3}}}} \right]} \right\}\left( { - {J^0}} \right) \nonumber\\
&+&72\mu _1^4{s^2}{\partial _i}\left\{ {\left[ {\frac{{{A_5}}}{{{{\left( {{\nabla ^2} - M_1^2} \right)}^2}}} - {A_6}\frac{{{\nabla ^2}}}{{{{\left( {{\nabla ^2} - M_1^2} \right)}^2}}}} \right] + \left[ {\frac{{{A_7}}}{{{{\left( {{\nabla ^2} - M_2^2} \right)}^2}}} - {A_8}\frac{{{\nabla ^2}}}{{{{\left( {{\nabla ^2} - M_2^2} \right)}^2}}}} \right]} \right\}\left( { - {J^0}} \right) \nonumber\\
&+&72\mu _1^4{s^2}{\partial _i}\left\{ {\left[ {\frac{{{A_9}}}{{\left( {{\nabla ^2} - M_1^2} \right)}} - {A_{10}}\frac{{{\nabla ^2}}}{{\left( {{\nabla ^2} - M_1^2} \right)}}} \right] - \left[ {\frac{{{A_9}}}{{\left( {{\nabla ^2} - M_2^2} \right)}} - {A_{10}}\frac{{{\nabla ^2}}}{{\left( {{\nabla ^2} - M_2^2} \right)}}} \right]} \right\}\left( { - {J^0}} \right),  \label{Sener80}
\end{eqnarray}
where $M_1^2 = \frac{1}{{2P}}\left( {Q + \sqrt {{Q^2} - 4PR} } \right)$ and $M_2^2 = \frac{1}{{2P}}\left( {Q - \sqrt {{Q^2} - 4PR} } \right)$, whereas the constants, $A_{1}-A_{10}$, are defined in the Appendix A.

For ${J^0}\left( {\bf x} \right) = q{\delta ^{\left( 2 \right)}}\left( {\bf x} \right)$, expression (\ref{Sener80}) becomes
\begin{eqnarray}
{E_i}({\bf x}) &=& \frac{q}{{\sqrt {{Q^2} - 4PR} }}{\partial _i}\left\{ {{\nabla ^2}{G_1}\left( {\bf x} \right) - {\nabla ^2}{G_2}\left( {\bf x} \right)} \right\} - \frac{{qD}}{{\sqrt {{Q^2} - 4PR} }}{\partial _i}\left\{ {{G_1}\left( {\bf x} \right) - {G_2}\left( {\bf x} \right)} \right\} \nonumber\\
& +& \frac{{qm_{\cal F}^4}}{{\sqrt {{Q^2} - 4PR} }}{\partial _i}\left\{ {\frac{{{G_1}\left( {\bf x} \right)}}{{{\nabla ^2}}} - \frac{{{G_2}\left( {\bf x} \right)}}{{{\nabla ^2}}}} \right\} \nonumber\\
&+& 72\mu _1^2{s^2}{\partial _i}\left\{ {\left[ {{A_1}{G_3}\left( {\bf x} \right) - {A_2}{\nabla ^2}{G_3}\left( {\bf x} \right)} \right] - \left[ {{A_3}{G_4}\left( {\bf x} \right) - {A_4}{\nabla ^2}{G_4}\left( {\bf x} \right)} \right]} \right\} \nonumber\\
&+& 72\mu _1^2{s^2}{\partial _i}\left\{ {\left[ {{A_5}{G_5}\left( {\bf x} \right) - {A_6}{\nabla ^2}{G_5}\left( {\bf x} \right)} \right] - \left[ {{A_7}{G_6}\left( {\bf x} \right) - {A_8}{\nabla ^2}{G_6}\left( {\bf x} \right)} \right]} \right\} \nonumber\\
&+& 72\mu _1^2{s^2}{\partial _i}\left\{ {\left[ {{A_9}{G_1}\left( {\bf x} \right) - {A_{10}}{\nabla ^2}{G_1}\left( {\bf x} \right)} \right] - \left[ {{A_9}{G_2}\left( {\bf x} \right) - {A_{10}}{\nabla ^2}{G_2}\left( {\bf x} \right)} \right]} \right\}. \label{Sener85}
\end{eqnarray}
To get last expression we have used: 
\begin{equation}
{G_1}\left( {\bf x} \right) =  - \frac{{{\delta ^{\left( 2 \right)}}\left( {\bf x} \right)}}{{{\nabla ^2} - M_1^2}} = \frac{1}{{2\pi }}{K_0}\left( {{M_1}|{\bf x}|} \right), \label{Sener85a}
\end{equation}
\begin{equation}
{G_2}\left( {\bf x} \right) =  - \frac{{{\delta ^{\left( 2 \right)}}\left( {\bf x} \right)}}{{{\nabla ^2} - M_2^2}} = \frac{1}{{2\pi }}{K_0}\left( {{M_2}|{\bf x}|} \right), \label{Sener85b}
\end{equation}
\begin{equation}
{G_3}\left( {\bf x} \right) =  - \frac{{{\delta ^{\left( 2 \right)}}\left( {\bf x} \right)}}{{{{\left( {{\nabla ^2} - M_1^2} \right)}^3}}} = \frac{1}{{4\pi }}\frac{{|{\bf x}{|^2}}}{{12M_1^2}}{K_{ - 2}}\left( {{M_1}|{\bf x}|} \right), \label{Sener85c}
\end{equation}
\begin{equation}
{G_4}\left( {\bf x} \right) =  - \frac{{{\delta ^{\left( 2 \right)}}\left( {\bf x} \right)}}{{{{\left( {{\nabla ^2} - M_2^2} \right)}^3}}} = \frac{1}{{4\pi }}\frac{{|{\bf x}{|^2}}}{{12M_2^2}}{K_{ - 2}}\left( {{M_2}|{\bf x}|} \right),\label{Sener85d}
\end{equation}
\begin{equation}
 {G_5}\left( {\bf x} \right) =  - \frac{{{\delta ^{\left( 2 \right)}}\left( {\bf x} \right)}}{{{{\left( {{\nabla ^2} - M_1^2} \right)}^2}}} =  - \frac{1}{{4\pi }}\frac{{|{\bf x}|}}{{{M_1}}}{K_{ - 1}}\left( {{M_1}|{\bf x}|} \right), \label{Sener85d}
\end{equation} 
\begin{equation}
{G_6}\left( {\bf x} \right) =  - \frac{{{\delta ^{\left( 2 \right)}}\left( {\bf x} \right)}}{{{{\left( {{\nabla ^2} - M_2^2} \right)}^2}}} =  - \frac{1}{{4\pi }}\frac{{|{\bf x}|}}{{{M_2}}}{K_{ - 1}}\left( {{M_2}|{\bf x}|} \right), \label{Sener85e}
\end{equation}
where $K_0$, $K_{-1}$ and $K_{-2}$ are modified Bessel functions. We also recall that ${K_{ - 1}}\left( z \right) =  - \left( {{K_1}\left( z \right) + 2\frac{d}{{dz}}{K_0}\left( z \right)} \right)$ and ${K_{ - 2}}\left( z \right) =  - \left( {{K_0}\left( z \right) - 2\frac{d}{{dz}}{K_1}\left( z \right) - 4\frac{{{d^2}}}{{d{z^2}}}{K_0}\left( z \right)} \right)$.
 
Let us also mention here that  \cite{Pato3}:
\begin{equation}
\frac{{{G_1}\left( {\bf x} \right)}}{{{\nabla ^2}}} = \frac{{|{\bf x}|}}{{4{M_1}}}, \label{Sener85f}
\end{equation} 
and 
\begin{equation}
\frac{{{G_2}\left( {\bf x} \right)}}{{{\nabla ^2}}} = \frac{{|{\bf x}|}}{{4{M_2}}}.\label{Sener85g}
\end{equation}

With this at hand, we now turn our attention to the calculation of the energy interaction between static point-like sources, by using the gauge-invariant but path-dependent variables formalism. To this end, we start by considering the expression \cite{Pato1}
\begin{equation}
V \equiv q\left( {{{\cal A}_0}\left( {\bf 0} \right) - {{\cal A}_0}\left( {\bf y} \right)} \right), \label{Sener90}
\end{equation}
where the physical scalar potential is given by
\begin{equation}
{{\cal A}_0}\left( {\bf x} \right) = \int_0^1 {d\lambda } {x^i}{E_i}\left( {\lambda {\bf x}} \right), \label{Sener95}
\end{equation}
and $i=1,2$. This follows from the vector gauge-invariant field expression \cite{Pato1}
\begin{equation}
{{\cal A}_\mu }\left( x \right) \equiv {A_\mu }\left( x \right) + {\partial _\mu }\left( { - \int_\xi ^x {d{z^\mu }{A_\mu }\left( z \right)} } \right), \label{Sener100}
\end{equation}
where the line integral is along a space-like path from $\xi$ to$x$, on a fixed time slice. it should be further noted that these variables (\ref{Sener100}) commute with the sole first class constraint (Gauss law), corroborating that these fields are physical variables.

With the aid of equation (\ref{Sener85}), equation (\ref{Sener95}) becomes
\begin{eqnarray}
{{\cal A}_0}({\bf x}) &=& \frac{q}{{\sqrt {{Q^2} - 4PR} }}\left\{ {{\nabla ^2}{G_1}\left( {\bf x} \right) - {\nabla ^2}{G_2}\left( {\bf x} \right)} \right\} - \frac{{qD}}{{\sqrt {{Q^2} - 4PR} }}\left\{ {{G_1}\left( {\bf x} \right) - {G_2}\left( {\bf x} \right)} \right\} \nonumber\\
& +& \frac{{qm_{\cal F}^4}}{{\sqrt {{Q^2} - 4PR} }}\left\{ {\frac{{{G_1}\left( {\bf x} \right)}}{{{\nabla ^2}}} - \frac{{{G_2}\left( {\bf x} \right)}}{{{\nabla ^2}}}} \right\} \nonumber\\
&+& 72\mu _1^2{s^2}q\left\{ {\left[ {{A_1}{G_3}\left( {\bf x} \right) - {A_2}{\nabla ^2}{G_3}\left( {\bf x} \right)} \right] - \left[ {{A_3}{G_4}\left( {\bf x} \right) - {A_4}{\nabla ^2}{G_4}\left( {\bf x} \right)} \right]} \right\} \nonumber\\
&+& 72\mu _1^2{s^2}q\left\{ {\left[ {{A_5}{G_5}\left( {\bf x} \right) - {A_6}{\nabla ^2}{G_5}\left( {\bf x} \right)} \right] - \left[ {{A_7}{G_6}\left( {\bf x} \right) - {A_8}{\nabla ^2}{G_6}\left( {\bf x} \right)} \right]} \right\} \nonumber\\
&+& 72\mu _1^2{s^2}q\left\{ {\left[ {{A_9}{G_1}\left( {\bf x} \right) - {A_{10}}{\nabla ^2}{G_1}\left( {\bf x} \right)} \right] - \left[ {{A_9}{G_2}\left( {\bf x} \right) - {A_{10}}{\nabla ^2}{G_2}\left( {\bf x} \right)} \right]} \right\}, \label{Sener105}
\end{eqnarray}
after subtracting the self-energy term.

From equations (\ref{Sener90}) and (\ref{Sener105}), the corresponding static potential for two opposite charges located at ${\bf 0}$ and ${\bf y}$ may be written as
\begin{eqnarray}
V &=&  - \frac{{{q^2}}}{{2\pi }}\frac{D}{{\sqrt {{Q^2} - 4PR} }}\left( {{K_0}\left( {{M_2}L} \right) - {K_0}\left( {{M_1}L} \right)} \right) + \frac{{{q^2}m_F^4}}{{\sqrt {{Q^2} - 4PR} }}\left( {\frac{1}{{{M_2}}} - \frac{1}{{{M_1}}}} \right)L \nonumber\\
 &-& \frac{{{q^2}}}{{2\pi }}\frac{1}{{\sqrt {{Q^2} - 4PR} }}\left( {{\nabla ^2}{K_0}\left( {{M_1}L} \right) - {\nabla ^2}{K_0}\left( {{M_2}L} \right)} \right) \nonumber\\
&-& 72\mu _1^2{s^2}q^2\left\{ {\left[ {{A_1}{G_3}\left( {\bf y} \right) - {A_2}{\nabla ^2}{G_3}\left( {\bf y} \right)} \right] - \left[ {{A_3}{G_4}\left( {\bf y} \right) - {A_4}{\nabla ^2}{G_4}\left( {\bf y} \right)} \right]} \right\} \nonumber\\
&+& 72\mu _1^2{s^2}q^2\left\{ {\left[ {{A_7}{G_6}\left( {\bf y} \right) - {A_8}{\nabla ^2}{G_6}\left( {\bf y} \right)} \right] - \left[ {{A_5}{G_5}\left( {\bf y} \right) - {A_6}{\nabla ^2}{G_5}\left( {\bf y} \right)} \right]} \right\} \nonumber\\
&-& 72\mu _1^2{s^2}q^2\left\{ {\left[ {{A_9}{G_1}\left( {\bf y} \right) - {A_{10}}{\nabla ^2}{G_1}\left( {\bf y} \right)} \right] - \left[ {{A_9}{G_2}\left( {\bf y} \right) - {A_{10}}{\nabla ^2}{G_2}\left( {\bf y} \right)} \right]} \right\}, \label{Sener110}
\end{eqnarray}
where $L \equiv  |{\bf y}|$.

It is worthy noting here that the three first terms on the right hand side of expression (\ref{Sener110}) are at leading order in the coupling constant. In fact, this part of the potential displays a screening part, encoded in the Bessel functions and their derivatives, and the linear confining potential. As expected, this confinement disappears when $m \to 0$ (${m_{\cal F}} \to 0$). Interestingly, it is observed that the two first terms on the right hand side of expression (\ref{Sener110}) is exactly the result obtained for $D=3$ models of antisymmetric tensor fields that results from the condensation of topological defects, as a consequence of the Julia-Thoulousse mechanism \cite{Pato3}. As well as, by the condensation of charged scalars in $D=3$ dimensions \cite{Pato4}. We should also mention that in higher order in the coupling constant, a confining potential appears by means of $G_5$ and $G_6$ functions. Finally, we highlight the confining logarithm behavior displayed by the potential $V$: For ${M_1}L \ll 1, {M_2}L \ll 1$ yield $\ln ({M_1}L), \ln ({M_2}L)$ terms,
which is compatible with the energy for the vortex-antivortex interaction calculated in the paper of Ref. \cite{Shapos2}.

\subsection{Chern-Simons effective theory II (integrating the ${\cal A}_\mu$-field)}

We now wish to repeat what we have done in the previous Subsection when the ${\cal A}$-field shall be eliminated in favor of the ${\cal Z}$-field in equation (\ref{CSm45}). Going back to the Lagrangian density of equation (\ref{CSm45}), it is worthy to notice that the ${\cal A}_\mu$-field appears only through its field-strength, ${\cal F}_{\mu\nu}$, and the latter is only present quadratically and linearly. This means that ${\cal F}_{\mu\nu}$ appears as an auxiliary field. However, care must be taken in eliminating it, for it satisfies the constraint ${\partial _\mu }{\tilde {\cal {\cal F}}^{\mu}} = 0$, ${\tilde {\cal {\cal F}}^\mu }$ standing for its dual ${\tilde {\cal {\cal F}}^\mu } = \frac{1}{2}{\varepsilon ^{\mu \nu \kappa }}{{\cal {\cal F}}_{\nu \kappa }}$.

Before going on to eliminate the ${\cal A}_\mu$-field, we stress that, contrary to what we have done in Section III.A (where we have integrated over the massive mode, ${\cal Z}_\mu$, to formulate an effective model for ${\cal A}_\mu$, valid in a scale distance above the Compton wavelength of the ${\cal Z}$-particle), here, by integrating out the ${\cal A}_{\mu}$-field, which is massless, the idea is not the same as in the previous situation: we are
not writing down an effective physical model for  ${\cal Z}_\mu$. We are actually summing
up the effects of ${\cal A}_\mu$  ( because it only appears at most quadratically). So, the
procedures of integrating over ${\cal Z}_\mu$ and ${\cal A}_\mu$ are based on different lines of arguments.

Once we have made this remark, let us go ahead and eliminate ${\cal A}_\mu$ by actually eliminating ${\cal F}_{\mu\nu}$: by rewriting ${\cal F}_{\mu\nu}$ in terms of $\tilde {\cal F}^{\mu}$ and introducing a Lagrange multiplier field, $\chi$, to take into account the constraint (actually, the Bianchi identity) on
$\tilde {\cal F}^{\mu}$, we can carry out a chain of field reshufflings to finally arrive at the Lagrangian
\begin{equation}
{\cal L} =  - \frac{1}{4}{\cal S}_{\mu \nu }^2 + \frac{1}{2}{\mu _2}{\varepsilon ^{\mu \nu \kappa }}{{\cal S}_{\mu \nu }}{S_\kappa } + \frac{1}{2}\mu _1^2S_\mu ^2 - \frac{1}{2}{\cal H}_\mu ^2 + {\cal L}\left( {\hat \phi ,{{\hat \phi }^ * }} \right), \label{Sener115}
\end{equation}
where ${{\cal S}_\mu } = {{\cal Z}_\mu } - \frac{1}{{2{\mu _1}}}{\partial _\mu }\chi$, ${{\cal H}_\mu } = {\tilde {\cal F}_\mu } - 2{\mu _1}{{\cal S}_\mu }$, $\phi  = \hat \phi {e^{\frac{{ie}}{{2{\mu _1}}}\chi }}$ and ${D_\mu } = {\partial _\mu }-ie{{\cal S}_\mu }$.

We notice that the Lagrange multiplier field plays the role of a compensating field for the U(1)-symmetry associated to ${\cal Z}_\mu$, so that ${\cal S}_\mu$ is gauge-invariant, though it exhibits a Proca-type mass term along with its topological mass. The Lagrange multiplier field sets up a Stuckelberg formulation for the ${\cal S}_\mu$-field. The charged scalar undergoes a phase redefinition through the $\chi$-field, which ensures that it couples minimally to ${\cal S}_\mu$, and the redefined ${{\cal H}_\mu }$-field, which only appears algebraically and is completely decoupled, may be immediately eliminated by its Euler-Lagrange field equation.

After these field redefinitions have been implemented, we can go on with the potential ${\cal S}_\mu$ and the redefined scalar field to consider the phase where the latter spontaneously breaks the Abelian symmetry. We adopt in the sequel the unitary gauge choice.

Again, by splitting the $\hat \phi$-field into a background value and a dynamical part
\begin{equation}
\hat \phi  = {\hat \phi ^ * } = {\phi _0} + \frac{1}{{\sqrt 2 }}\eta \left( x \right), \label{Sener115a}
\end{equation}
where ${\phi _0} = \sqrt {\frac{2}{\lambda }} |m|$, we expand the Lagrangian up to quadratic terms in the fluctuations. Thus, the corresponding effective Lagrangian density reads 
\begin{equation}
{\cal L} =  - \frac{1}{4}{{\cal S}_{\mu \nu }}{{\cal S}^{\mu \nu }} + \frac{{{M^2}}}{2}{{\cal S}_\mu }{{\cal S}^\mu } + \frac{{{\mu _2}}}{2}{\varepsilon ^{\mu \nu \alpha }}{{\cal S}_{\mu \nu }}{{\cal S}_\alpha } - {\cal S}_{0}J^{0}, \label{Sener120}
\end{equation}
where $\frac{{{M^2}}}{2} \equiv {e^2}\phi _0^2 + 2\mu _1^2$ and $J^0$ is an external source.

This new effective theory provide us with a suitable starting point to study the interaction energy. Nevertheless, one can further observe that before proceeding with the determination of this energy, we need to restore the gauge invariance in equation (\ref{Sener120}). Making use of standard techniques for constrained systems, we find that equation (\ref{Sener120}) reduces to
\begin{equation}
{\cal L} =  - \frac{1}{4}{{\cal S}_{\mu \nu }}\left( {1 + \frac{{{M^2}}}{\Delta }} \right){{\cal S}^{\mu \nu }} + {\mu _2}{\varepsilon ^{\alpha \mu \nu }}{{\cal S}_\alpha }{\partial _\mu }{{\cal S}_\nu } - {\cal S}_{0}J^{0}. \label{Sener125}
\end{equation}
Notice that, for notational convenience, we have maintained $\Delta$ in equation (\ref{Sener125}), but it should be borne in mind that we are considering the static case.

With this in hand, the canonical momenta are ${\Pi ^\mu } =  - \left( {1 + \frac{{{M^2}}}{\Delta }} \right){{\cal S}^{0\mu }} + {\mu _2}{\varepsilon ^{\alpha 0\mu }}{{\cal S}_\alpha }$, and one immediately identifies the primary constraint ${\Pi ^0} \approx 0$. Whereas the remaining non-zero momenta are ${\Pi ^i} = \left( {1 + \frac{{{M^2}}}{\Delta }} \right){{\cal S}^{i0}} + {\mu _2}{\varepsilon ^{ij}}{{\cal S}_j}$. Now, the canonical Hamiltonian of this theory can be worked out as before and is given by
\begin{equation}
{H_C} \approx \int {{d^2}x} \left\{ { - {{\cal S}_0}\left( {{\partial _i}{\Pi ^i} + {\mu _2}{\varepsilon ^{ij}}{\partial _i}{{\cal S}_j -J^0}} \right) + \frac{1}{2}{E^i}\left( {1 + \frac{{{M^2}}}{\Delta }} \right){E^i} + \frac{1}{2}B\left( {1 + \frac{{{M^2}}}{\Delta }} \right)B} \right\}.  \label{Sener130}
\end{equation}
Once again, requiring the primary constraint ${\Pi ^0}$ to be preserved in time yields the secondary constraint (Gauss's law) ${\Gamma _1} \equiv {\partial _i}{\Pi ^i} + {\mu _2}{\varepsilon ^{ij}}{\partial _i}{{\cal S}_j} - {J^0} \approx 0    $. By proceeding in the same way as before, the Hamiltonian turns out to be
\begin{equation}
{H} = \int {{d^2}x} \left\{ {  {w(x)}\left( {{\partial _i}{\Pi ^i} + {\mu _2}{\varepsilon ^{ij}}{\partial _i}{{\cal S}_j - J^0}} \right) + \frac{1}{2}{E^i}\left( {1 + \frac{{{M^2}}}{\Delta }} \right){E^i} + \frac{1}{2}B\left( {1 + \frac{{{M^2}}}{\Delta }} \right)B} \right\}.  \label{Sener135}
\end{equation}

Since our goal is to compute the static potential for the theory under consideration, we shall adopt the same gauge-fixing condition that was used in our preceding calculation. In view of this situation, we now proceed to write the Dirac brackets in terms of the magnetic and electric fields as
\begin{equation}
{\left\{ {{E_i}\left( x \right),B\left( y \right)} \right\}^ * } =  - {\left( {1 + \frac{{{M^2}}}{\Delta }} \right)^{ - 1}}{\varepsilon _{ij}}\partial _x^j{\delta ^{\left( 2 \right)}}\left( {x - y} \right), \label{Sener140}
\end{equation}

\begin{equation}
{\left\{ {B\left( x \right),B\left( y \right)} \right\}^ * } = 0, \label{Sener145}
\end{equation}

\begin{equation}
{\left\{ {{E_i}\left( x \right),{E_j}\left( y \right)} \right\}^ * } =  - 2{\mu _2}{\left( {1 + \frac{{{M^2}}}{\Delta }} \right)^{ - 2}}{\varepsilon _{ij}}{\delta ^{\left( 2 \right)}}\left( {x - y} \right), \label{Sener150}
\end{equation}

It gives rise to the following equations of motion for ${E_i}$ and $B$ fields:
\begin{equation}
{\dot E_i}\left( x \right) =  - 2{\mu _2}{\left( {1 + \frac{{{M^2}}}{\Delta }} \right)^{ - 2}}{\varepsilon _{ij}}{E_j}\left( x \right) + {\left( {1 + \frac{{{M^2}}}{\Delta }} \right)^{ - 1}}{\varepsilon _{ij}}{\partial _j}B\left( x \right), \label{Sener155}
\end{equation}

\begin{equation}
\dot B\left( x \right) =  - {\left( {1 + \frac{{{M^2}}}{\Delta }} \right)^{ - 1}}{\varepsilon _{ij}}{\partial _i}{E_j}\left( x \right). \label{Sener160}
\end{equation}

Note that Gauss law for the present theory reads
\begin{equation}
\left( {1 + \frac{{{M^2}}}{\Delta }} \right){\partial _i}{E^i} + 2{\mu _2}B - {J^0} = 0, \label{Sener165}
\end{equation}

As before, we shall consider static fields. Therefore, the electric field assumes the form
\begin{equation}
{E_i} = \frac{1}{{\mu \sqrt {4{M^2} - {\mu ^2}} }}{\partial _i}\left\{ {\left( {\frac{{{\nabla ^2}}}{{{\nabla ^2} - M_1^2}} - \frac{{{\nabla ^2}}}{{{\nabla ^2} - M_2^2}}} \right) - {M^2}\left( {\frac{1}{{{\nabla ^2} - M_1^2}} - \frac{1}{{{\nabla ^2} - M_2^2}}} \right)} \right\}\left( { - {J^0}} \right),  \label{Sener170}
\end{equation}
with $\mu  \equiv 2{\mu _2}$. Here $M_1^2 = \frac{1}{2}\left[ {2{M^2} + {\mu ^2} + \mu \sqrt {4{M^2} + {\mu ^2}} } \right]$ and $M_2^2 = \frac{1}{2}\left[ {2{M^2} + {\mu ^2} - \mu \sqrt {4{M^2} + {\mu ^2}} } \right]$.

We now have all the information required to compute the potential energy for static charges in this theory. Thus, by employing equation (\ref{Sener95}), for ${J^0}\left( {\bf x} \right) = q{\delta ^{\left( 2 \right)}}\left( {\bf x} \right)$, the gauge-invariant scalar potential takes the form
\begin{equation}
{{\cal A}_0}\left( {\bf x} \right) = \frac{q}{{\mu \sqrt {{M^2} - {\mu ^2}} }}\left\{ {\left( {{\nabla ^2}{G_1}\left( {\bf x} \right) - {\nabla ^2}{G_2}\left( {\bf x} \right)} \right) - {M^2}\left( {{G_1}\left( {\bf x} \right) - {G_2}\left( {\bf x} \right)} \right)} \right\}. \label{Sener175}
\end{equation}

Finally, making use of equation (\ref{Sener90}), the potential energy for a pair of static point-like opposite charges at ${\bf 0}$ and ${\bf y}$, becomes
\begin{equation}
V =  - \frac{{{q^2}}}{{2\pi }}\frac{{{M^2}}}{{\mu \sqrt {{M^2} - {\mu ^2}} }}\left( {{K_0}\left( {{M_1}L} \right) - {K_0}\left( {{M_2}L} \right)} \right) + \frac{{{q^2}}}{{2\pi }}\frac{1}{{\mu \sqrt {{M^2} - {\mu ^2}} }}\left( {{\nabla ^2}{K_0}\left( {{M_1}L} \right) - {\nabla ^2}{K_0}\left( {{M_2}L} \right)} \right).  \label{Sener180}
\end{equation}

We immediately see that, unexpectedly, the confining potential between static charges vanishes in this case. In other words, this effective theory describes an exactly screening phase.

\section{Final Remarks}

In summary, within the gauge-invariant but path-dependent variables formalism, we have considered the confinement versus screening issue for a recently proposed  topologically massive $U{\left( 1 \right)_{\cal W}} \times U{(1)_{\cal Y}}$ Chern-Simons-Higgs theory in $2+1$ dimensions. Once again, a correct identification of physical degrees of freedom has been fundamental for understanding the physics hidden in gauge theories. It was shown, that in the case of an effective Lagrangian in terms of the $\cal A$ field the interaction energy displays a screening part, encoded by Bessel functions, and a linear confining potential. Incidentally, the above static potential profile is similar to that encountered in both Maxwell-like three-dimensional models induced by the condensation of topological defects \cite{Pato3} and by the condensation of charged scalars in $D=3$ dimensions \cite{Pato4}. In this way, we may establish a new connection among diverse models as well as exploiting this equivalence in explicit calculations. However, in the case of an effective Lagrangian in terms of the $\cal Z$ field, the surprising result is that the theory describes an exactly screening phase. Actually, contrary to the case of the ${\cal A}_\mu$-field effective model, in the situation of Section III.B, we have completely eliminated the effects of the massless mode, ${\cal A}_\mu$, so that the ${\cal Z}_\mu$-field based model that comes out is genuinely massive, and it should only display screening. Contrary, in Section III.A, we keep the massless field ${\cal A}_\mu$ and add up contributions which arise from the effects of integrating out ${\cal Z}_\mu$, which is valid for energies much below its mass. Therefore, we should not expect to loose the confining effect typical of planar massless modes. We believe this should be the way to understand why there appears no confinement in the model stemming from the elimination of the ${\cal A}_\mu$-field, while the first situation exhibits both confinement and screening.

\section{ACKNOWLEDGMENTS}
P. G. was partially supported by Fondecyt (Chile) grant 1130426, DGIP (UTFSM) internal project USM 111458. P. G. also wishes to thank the Field Theory Group of the CBPF for hospitality.

\section{Appendix A}

Below, we collect the constants $A_{1}-A_{10}$:
\begin{equation}
{A_1} =  - \frac{R}{P}\frac{1}{{{{\left( {{Q^2} - 4PR} \right)}^{{\raise0.5ex\hbox{$\scriptstyle 3$}
\kern-0.1em/\kern-0.15em
\lower0.25ex\hbox{$\scriptstyle 2$}}}}}}\frac{1}{{M_2^2}}\left( {D - \frac{{m_{\cal F}^4}}{{M_1^2}}} \right), \label{app05}
\end{equation}
\begin{equation}
{A_2} =  - \frac{{M_1^2}}{{{{\left( {{Q^2} - 4PR} \right)}^{{\raise0.5ex\hbox{$\scriptstyle 3$}
\kern-0.1em/\kern-0.15em
\lower0.25ex\hbox{$\scriptstyle 2$}}}}}}, \label{app10}
\end{equation}
\begin{equation}
{A_3} =  - \frac{R}{P}\frac{1}{{{{\left( {{Q^2} - 4PR} \right)}^{{\raise0.5ex\hbox{$\scriptstyle 3$}
\kern-0.1em/\kern-0.15em
\lower0.25ex\hbox{$\scriptstyle 2$}}}}}}\frac{1}{{M_1^2}}\left( {D - \frac{{m_{\cal F}^4}}{{M_2^2}}} \right), \label{app15}
\end{equation}
\begin{equation}
{A_4} =  - \frac{{M_2^2}}{{{{\left( {{Q^2} - 4PR} \right)}^{{\raise0.5ex\hbox{$\scriptstyle 3$}
\kern-0.1em/\kern-0.15em
\lower0.25ex\hbox{$\scriptstyle 2$}}}}}}, \label{app20}
\end{equation}
\begin{equation}
{A_5} = \frac{R}{{{{\left( {{Q^2} - 4PR} \right)}^2}}}\left\{ {\frac{2}{{M_2^2}}\left( {D - \frac{{m_{\cal F}^4}}{{M_1^2}}} \right) + \frac{1}{{M_1^2}}\left( {D - \frac{{m_{\cal F}^4}}{{M_2^2}}} \right)} \right\}, \label{app25}
\end{equation}
\begin{equation}
{A_6} = \frac{P}{{{{\left( {{Q^2} - 4PR} \right)}^2}}}\left( {2M_1^2 + M_2^2} \right), \label{app30}
\end{equation}
\begin{equation}
{A_7} = \frac{R}{{{{\left( {{Q^2} - 4PR} \right)}^2}}}\left\{ {\frac{1}{{M_2^2}}\left( {D - \frac{{m_{\cal F}^4}}{{M_1^2}}} \right) + \frac{2}{{M_1^2}}\left( {D - \frac{{m_{\cal F}^4}}{{M_2^2}}} \right)} \right\}, \label{app35}
\end{equation}
\begin{equation}
{A_8} = \frac{P}{{{{\left( {{Q^2} - 4PR} \right)}^2}}}\left( {2M_2^2 + M_1^2} \right), \label{app40}
\end{equation}
\begin{equation}
{A_9} =  - \frac{{3PQD}}{{{{\left( {{Q^2} - 4PR} \right)}^{{\raise0.5ex\hbox{$\scriptstyle 5$}
\kern-0.1em/\kern-0.15em
\lower0.25ex\hbox{$\scriptstyle 2$}}}}}}, \label{app40}
\end{equation}
\begin{equation}
{A_{10}} =  - \frac{{3PQ}}{{{{\left( {{Q^2} - 4PR} \right)}^{{\raise0.5ex\hbox{$\scriptstyle 5$}
\kern-0.1em/\kern-0.15em
\lower0.25ex\hbox{$\scriptstyle 2$}}}}}}. \label{app45}
\end{equation}

\end{document}